\documentstyle[aps,multicol,epsf,epsfig]{revtex}

\begin{document}

\draft
%\tightenlines
 
\title{Long-range electrostatic interactions between like-charged
colloids: steric and confinement effects}

\author{Emmanuel Trizac$^1$ and Jean-Luc Raimbault$^2$}

\address{
$^1$ Laboratoire de Physique Th\'eorique\cite{umr1}, B\^atiment 210, Universit\'e 
de Paris-Sud, 91405 Orsay Cedex, France\\
E-mail: Emmanuel.Trizac@th.u-psud.fr\\
$^2$ Centre de Recherches sur la Mati\`ere Divis\'ee\cite{umr2}, 
Universit\'e d'Orl\'eans,
45071 Orl\'eans C\'edex, France\\
E-mail: raimbau@cnrs-orleans.fr
}
 
\date{\today}

\maketitle
\begin{abstract}
Within the framework of a Modified Poisson-Boltzmann
theory accounting for steric effects of microions, 
we prove analytically that the effective pair interactions
between like-charge colloids immersed in a confined electrolyte 
are repulsive. Our approach encompasses and extends previously known results 
to the case of complete confinement, and further incorporates
the finite size of the microions which is absent in the standard Poisson-Boltzmann
theory.
\end{abstract}


%\begin{multicols}{2}

Recent experiments show convincingly the existence of long-range electrostatic
attractions between like-charge colloids immersed in an electrolyte, in particular
in the vicinity of a charged wall or when the particles are confined in a slit
\cite{Larsen,Grier}. These striking observations are inconsistent with the
well established theory of Derjaguin, Landau, Verwey and Overbeek
(DLVO) \cite{dlvo} and constitute an important controversy in colloid 
science. Inasmuch as direct measurements of the effective pair
interactions in the {\em bulk} of a suspension confirm the validity
of DLVO theory with repulsive interactions \cite{Vonder}, confinement
effects are thought to play a major role and have been included in 
numerical \cite{Bowen} and theoretical \cite{Goulding}
investigations of the electrostatic forces between colloidal spheres.
These studies reported a range of inter-particle distances for which the 
computed forces were attractive. However, Neu \cite{Neu} rigorously proved that the 
Poisson-Boltzmann (PB) model considered in \cite{Bowen} necessarily leads
to repulsion. His proof applies  within PB theory (which forms the basis
of the DLVO potential) in the specific case of Dirichlet boundary 
conditions (constant electrostatic potential $\psi$ on the confining surface, 
the experimental relevance of which is dubious).
Sader and Chan extended the PB argument to a broader class of boundary conditions
\cite{Sader}.

The PB theory is a continuum mean-field approach neglecting statistical
correlations between microions, that are assumed to be point-like. It is
known to grossly overestimate counter-ion concentrations close to a charged
surface (eg that of a polyion). A Modified Poisson-Boltz\-mann (MPB)
theory including steric effects was put forward to overcome this
shortcoming \cite{Eigen,Iglic,Borukhov,Gouldingbis}. 
This approach retains the simplicity of the original PB theory and is more
tractable than the other attempts to improve upon PB by inclusion of steric 
repulsion \cite{Stern,Kje}.

In this article, we consider the MPB model in the situation of 
a mixture of microions (with overall global electroneutrality). 
Within this framework, we analyze the
interactions between a pair of like-charge polyions
immersed in a confined electrolyte of permittivity $\varepsilon$. 
The colloids may be of arbitrary shape and
the confining region ${\cal R}$ is a cylinder of arbitrary cross-section.
As in \cite{Neu,Sader}, the only requirement is
that the electrostatic potential $\psi$ possesses mirror symmetry with 
respect to the mid plane $z=0$ between the colloids.
Unlike in \cite{Neu,Sader} where ${\cal R}$ is of infinite lateral extension,
we allow the confining region to be of finite volume (situation hereafter
refered as finite confinement, which can be that 
of a pore or of a closed Wigner-Seitz cell). 
The geometry considered here encompasses many cases of experimental
relevance, such as two colloids confined in a slit, in a charged
pore or in the vicinity of a wall \cite{Larsen,Grier,Goulding}.
We consider boundary conditions where the medium ${\cal R}'$
outside ${\cal R}$ 
is a dielectric continuum 
of permittivity $\varepsilon'$ with the possibility of a uniform
density of surface charges $\sigma$ on the boundary $\partial {\cal R}$
(as in \cite{Goulding}).
This class of boundary conditions is large, as for $\sigma=0$
the limit $\varepsilon' \to 0$ reduces to Neuman boundary conditions
with a vanishing normal electric field on $\partial {\cal R}$, whereas
$\varepsilon' \to \infty$ corresponds to the standard Dirichlet
condition on $\partial {\cal R}$. 
We show rigorously that the inclusion of excluded volume effects
within the MPB approach does not change the sign of the force which 
remains repulsive as in PB.
As the MPB gives back the PB theory in the well behaved limit where
steric effects disappear, our results also include those reported in \cite{Neu,Sader},
and extend them to the situation of finite confinement.

The MPB theory can be obtained from the free energy functional 
${\cal F} = {\cal F}_{\hbox{\scriptsize Coul}} +
{\cal F}_{\hbox{\scriptsize ent}}$ of the inhomogeneous
fluid of microions contained in ${\cal R}$. Given $N$ species $\alpha$
with a local density $c_{\alpha}(\bbox{r})$ and charge number $z_\alpha$,
the Coulombic energy contribution to ${\cal F}$ reads
\begin{equation}
{\cal F}_{\hbox{\scriptsize Coul}} = \frac{1}{2} \, \int_{\cal R} 
\rho_c(\bbox{r})\, G(\bbox{r}, \bbox{r}') \, \rho_c(\bbox{r}')\, d\bbox{r}
d\bbox{r}',
\label{eq:coul}
\end{equation}
where $\rho_c(\bbox{r}) = \sum_\alpha  z_\alpha e c_\alpha(\bbox{r})$ is the local
charge density and $G(\bbox{r}, \bbox{r}')$ is the Green's function 
inverting the Laplacian inside ${\cal R}$ with the required boundary conditions.
In the absence of confinement, $G$ reduces to the Coulomb potential
$1/[4 \pi\varepsilon |\bbox{r}-\bbox{r}'|]$. 
The entropic contribution is
\begin{equation}
{\cal F}_{\hbox{\scriptsize ent}} = kT \, \int_{\cal R} d \bbox{r}\,
\left\{\sum_\alpha c_\alpha \left[\ln\left(\Lambda_\alpha^3 c_\alpha\right)-1\right]
+\frac{1}{a^3} \left[1-\sum_\alpha a^3 c_\alpha\right]\ln\left(
1-\sum_\alpha a^3 c_\alpha\right) \right\},
\label{eq:ent}
\end{equation}
where $\Lambda_\alpha$ are irrelevant length scales and $kT \equiv \beta^{-1}$
is the thermal energy. For the sake of simplicity,
all types of microions are assumed to have the same
size $a$.
The last term in Eq. (\ref{eq:ent}) mimics the entropy of the solvent molecules
while the remaining ones are the ideal entropies of the mixture of co- and
counter-ions. In the limit where $a \to 0$, the steric correction disappears
and the classical PB expression
is recovered. The equilibrium density profiles
are those which minimize the free energy functional ${\cal F}$
subject to the constraint
that
\begin{equation}
\int_{\cal R} c_\alpha(\bbox{r})\, d\bbox{r}\,=\, N_\alpha
\label{eq:norm}
\end{equation}
with $N_\alpha$ the total number of ions $\alpha$ inside ${\cal R}$.
%\begin{equation}
%\frac{\delta {\cal F}}{\delta c_alpha(\bbox{r})} = \mu_\alpha,
%\end{equation}
%where $\mu_\alpha$ is the chemical potential of species $\alpha$. 
The minimization yields:
\begin{equation}
c_\alpha(\bbox{r}) = \frac{p_\alpha\, \exp(-\beta e z_\alpha \psi)}{1+
\sum_\alpha a^3 p_\alpha \exp(-\beta e z_\alpha \psi)},
\end{equation}
where 
$\psi(\bbox{r}) = \int_{\cal R} \rho_c(\bbox{r}') G(\bbox{r},\bbox{r}') \,d\bbox{r}'$.
Since the potential $\psi$ can be shifted by 
an arbitrary constant, the prefactors
$p_\alpha$ have individually 
no physical significance. In the limit where ${\cal R}$
is of infinite extension, $\psi$ is conveniently chosen to vanish far from 
the polyions and we get
$p_\alpha = c_\alpha^{\hbox{\scriptsize bulk}}
/(1-\sum_\alpha a^3 c_\alpha^{\hbox{\scriptsize bulk}})$. 
In the case of complete confinement, two experimental
situations can be distinguished: a) the solution is of fixed ionic
composition (canonical description)
and the prefactors $p_\alpha$ are determined by the normalization 
constraint (\ref{eq:norm}), or b) the dispersion is in osmotic equilibrium
with a salt reservoir, treated within MPB for consistency. We then have 
$p_\alpha = c_\alpha^r /(1-\sum_\alpha a^3 c_\alpha^r)$, where 
$c_\alpha^r$ denotes the concentration of species $\alpha$ in the reservoir
(subject to the electroneutrality constraint $\sum_\alpha z_\alpha \,c_\alpha^r=0$).
As expected, the density profiles tend to their PB counterparts as the 
excluded volume $a^3 \to 0$.

The force acting on a colloidal particle follows from integration of the stress
tensor $\bbox{\Pi}$ over the surface $S$ of the polyion (see Fig 1):
\begin{equation}
\bbox{F} = \oint_S \bbox{\Pi} \cdot \bbox{\hat n}\, dS,
\label{eq:forcedef}
\end{equation}
where $\bbox{\hat n}$ is the unit vector pointing outwards from 
the surface of integration. 
The stress tensor can be obtained by considering the mechanical equilibrium
condition of a fluid element of microions: the balance between the electric force
and the os\-mo\-tic constraint can be written
\begin{eqnarray}
&&-\bbox{\nabla} P \,+\, \rho_c \,\bbox{E} = \bbox{0} 
\label{eq:equil}\\
&\Leftrightarrow & \,\, \partial_\psi P = -\rho_c,
\end{eqnarray}
from which we deduce the expression of the osmotic pressure:
\begin{equation}
P(\psi) = \frac{kT}{a^3}\, \ln\left[1+\sum_\alpha a^3 p_\alpha 
\exp(-\beta e z_\alpha \psi)\right].
\label{eq:osmotic}
\end{equation}
Rewriting Eq. (\ref{eq:equil}) as $\bbox{\nabla}\cdot \bbox{\Pi} = \bbox{0}$
finally yields 
\begin{equation}
\bbox{\Pi}= -\left[ P(\psi) + \frac{1}{2} \,\bbox{D}\cdot\bbox{E} \right]
\bbox{I} + \,\bbox{D} \otimes\bbox{E},
\label{eq:ptensor}
\end{equation}
where $\bbox{E} = -\bbox{\nabla}\psi$ is the electrostatic field,
$\bbox{D}=\varepsilon \bbox{E}$ and $\bbox{I}$ denotes the identity tensor. 
It is worthwhile to mention that Poisson's equation can be written in the form
\begin{equation}
\bbox{\nabla}^2 \psi = \frac{1}{\varepsilon} \frac{\partial P}{\partial \psi}.
\label{eq:Poisson}
\end{equation}
As the mechanical equilibrium condition invoked above implies that 
$\bbox{\Pi}$ is divergence free, the surface of integration
in (\ref{eq:forcedef})
can be deformed to any surface $S'$ enclosing colloid $S$ only, as 
stressed in \cite{Neu,Sader}. It is convenient to choose
$S'= \Sigma_0 \cup  \Sigma \cup  \Sigma_L$ (see Fig. 1). The $z$-component
of the force $\bbox{F}$ then reads
\begin{equation}
 F_z = \int_{ \Sigma_0 \cup \Sigma_L} \bbox{\hat z}\cdot \bbox{\Pi} 
\cdot \bbox{\hat n} \,\,dS +
\int_{\Sigma} \bbox{\hat z}\cdot \bbox{\Pi}\cdot \bbox{\hat n}\,\, dS.
\label{eq:for} 
\end{equation}
On the cross sections $\Sigma_0$ and $\Sigma_L$,
$\bbox{\hat z}\cdot\bbox{\Pi}\cdot\bbox{\hat n} =
[D_zE_z   - \bbox{D}\cdot\bbox{E}/2 - P(\psi)]\,\bbox{\hat z}\cdot\bbox{\hat n}$, 
while
on the lateral surface $\Sigma$, 
$\bbox{\hat z}\cdot\bbox{\Pi}\cdot\bbox{\hat n} = -\sigma E_z + D'_n E'_z$.
In the previous expression,
primed symbols refer to the region ${\cal R'}$ outside ${\cal R}$  (with
permittivity $\varepsilon'$) 
and the dielectric boundary conditions have been used 
($D'_n - D_n = \sigma$ and $E_z = E'_z$).
The last term in Eq. (\ref{eq:for}) can be recast by
considering the divergence free tensor $\bbox{\Pi}'$ in ${\cal R}'$
obeying 
similar constitutive relations as (\ref{eq:ptensor}) and
(\ref{eq:osmotic}) (we allow the continuum in ${\cal R}'$ to 
contain an electrolyte solution; on the other hand, 
if no salt is present outside
${\cal R}$, the corresponding pressure 
$P(\psi')$ vanishes in (\ref{eq:osmotic})). We get
\begin{equation}
\int_{ \Sigma} D'_n E'_z \,dS = \int_{ \Sigma'_0  }\left[
\left( D'_zE'_z   -\frac{1}{2} {\bbox{ D'}}\cdot{\bbox{E'}} - 
P(\psi') \right)_{z=L} -
\left( D'_zE'_z  -\frac{1}{2} {\bbox{ D'}}\cdot{\bbox{ E'}} - 
P(\psi')  \right)_{z=0} \right]\,dxdy,
\label{eq:fsigma}
\end{equation}
where $\Sigma_0 \cup \Sigma'_0$ is the $Oxy$ plane.

Upon substitution of (\ref{eq:fsigma}) into (\ref{eq:for}), 
the axial force can be expressed in the form
\begin{eqnarray}
 F_z &=& \int_{Oxy} \left[ P({\cal \psi})_{z=0} - P({\cal \psi})_{z=L} \right]
dx dy
+\frac{1}{2} \int_{Oxy} \left[ ({\bbox{D}}\cdot {\bbox{E}})_{z=0} - 
({\bbox{D}}\cdot {\bbox{E}})_{z=L} \right] dx dy
\nonumber\\
&-&\, \sigma \int_{ \Sigma}  E_z \,dS \,+\,
\int_{Oxy} \left[ (D_z\, E_z)_{z=L} - (D_z \, E_z)_{z=0} \right] dx dy,
\label{eq:finterme}
\end{eqnarray}
where primes have been omitted for the part of $Oxy$ belonging to
${\cal R}'$ (ie $\Sigma'_0$): when unambiguous,
the same notations are hereafter used
for the fields inside and outside ${\cal R}$, with 
$\epsilon$ standing for the permittivity
$\varepsilon$ in ${\cal R}$ 
and for $\varepsilon'$ in ${\cal R'}$. 
In the last term of (\ref{eq:finterme}), $(E_z)_{z=0} = 0$
due to the mirror symmetry and $(E_z)_{z=L}$ vanishes
only in the limit $L\to\infty$ considered in \cite{Neu,Sader}.
Expression (\ref{eq:finterme}) is conveniently recast invoking the 
identity
$$\left[ ({\bbox{D}}\cdot {\bbox{E}})_{z=0} - 
({\bbox{D}}\cdot {\bbox{E}})_{z=L}  \right]=
\epsilon \left( {\bbox{E}}_{z=0} - {\bbox{E}}_{z=L}\right)^2 +
2 \,{\bbox{E}}_{z=L} \cdot \left( {\bbox{D}}_{z=0} - {\bbox{D}}_{z=L}\right)$$
and the relation
\begin{eqnarray}
\int_{Oxy}  {\bbox{E}}_{z=L} \cdot \left( {\bbox{D}}_{z=0} -
{\bbox{D}}_{z=L}\right) dx dy 
&=&  \int_{\partial  \Sigma} \left( D_n -D'_n \right)_{z=L} \left(  
\psi_{z=0} - \psi_{z=L}\right) d\ell 
\nonumber\\
&& - \int_{Oxy} \left(\psi_{z=0} - \psi_{z=L}\right) 
\epsilon \bbox{\nabla}^2 \psi_{z=L} \, dx dy 
\label{eq:interme1}\\
&=& \sigma \int_{ \Sigma}  E_z \,dS
- \int_{Oxy} \left(   \psi_{z=0} - \psi_{z=L} \right)
\frac{\partial P}{\partial \psi}( \psi_{z=L}) \,dx dy.
\label{eq:interme2}
\end{eqnarray}
Equation. (\ref{eq:interme1}) follows from a standard Green identity. 
The line integral in (\ref{eq:interme1})
has been reexpressed remembering the
dielectric boundary condition $D'_n-D_n=\sigma$ and that
$\int_{\partial  \Sigma} \left( \psi_{z=L} - \psi_{z=0}\right) d\ell =
\int_{ \Sigma}  E_z dS$.
Expression (\ref{eq:interme2}) was then obtained making use of 
Poisson's equation (\ref{eq:Poisson}).

Gathering results, we finally obtain:
\begin{eqnarray}
 F_z &=& \int_{Oxy} \left[ P(\psi)_{z=0} - P(\psi)_{z=L} 
- \left(   \psi_{z=0} - \psi_{z=L}\right) 
\frac{\partial P}{\partial \psi}( \psi_{z=L})
\right] dx dy \nonumber\\
&+& \frac{\epsilon}{2} \int_{Oxy} \left( {\bbox{E}}_{z=0} - 
{\bbox{E}}_{z=L}\right)^2 dx dy\nonumber\\
&+&  \epsilon \int_{Oxy} \left[ (E_z)_{z=L}\right]^2 dx dy.
\label{eq:final}
\end{eqnarray}
It can be checked that since the coefficients $p_\alpha$ are positive,
the osmotic pressure $P$ is a convex-up function of its argument $\psi$.
Consequently, the first integral in (\ref{eq:final}) is positive
from which we conclude that $F_z \geq 0$. This rigorous result holds
irrespective of the specific boundary conditions to be applied on the
colloids and is independent
of the sign of the surface charge $\sigma$.
The complete confinement of the electrolyte solution in ${\cal R}$
results in an enhanced repulsion with respect to the geometries 
considered in \cite{Neu,Sader} for which the last term of (\ref{eq:final})
vanishes. As stressed by Neu \cite{Neu}, the non convexity of the pressure
$P(\psi)$ is a necessary condition for attractive interactions.
Note that this result holds beyond PB or MPB theories, in fact in any
continuum description of the electrolyte solution relying on the local
density approximation: once such a theory
provides the functional dependence of the charge density $\rho_c$ 
on the electrostatic potential $\psi$ (or equivalently of the pressure
$P$ appearing in the stress tensor (\ref{eq:ptensor})), 
$\partial_\psi \rho_c \leq 0$
(or equivalently $\partial^2_\psi P \geq 0$)
is a sufficient condition for repulsive pair interactions.

While the numerical results reported in \cite{Bowen} appear erroneous, 
the reason for the attraction found in \cite{Goulding} could lie in
the effective charge assigned to the polyions to account for
the presence of tightly bound counterions and polyion-microion excluded volume. 
Our alternative approach
to incorporate steric effects gives rise to repulsion. A more definite
answer could be obtained by considering the generic density
functional of Biben {\it et al.} \cite{Biben} for electric double layers 
which allows to cope with the finite size of the microions 
(treated as charged hard spheres) and the
molecular nature of the solvent (considered to be a mixture of
dipolar hard spheres). Our results nevertheless show the robustness
of repulsive interactions within mean-field theories and may point
to the importance of correlated microion density fluctuations
(neglected by density functional theories) in interpreting
the experimental data.

The authors acknowledge useful discussions with D. Andelman, 
B. Jancovici, F. van Wijland, O. Martin and A. Barrat. 
JLR thanks the complex fluid 
group in CRMD (Orl\'eans) for stimulating interactions.

\newpage

\begin{center}
Figure 1 (Trizac and Raimbault)
\end{center}
%\vskip -1cm
\begin{figure}
$$\input{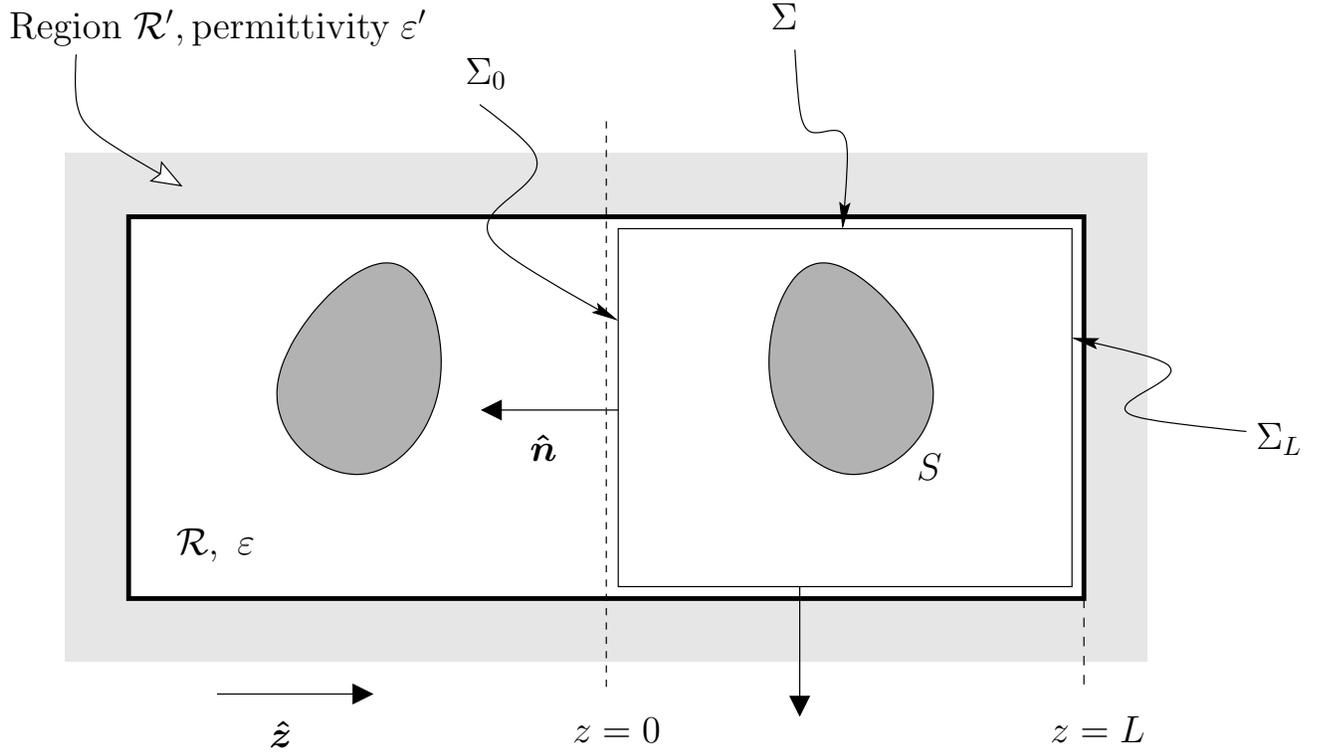}$$
\vskip 1cm
\caption{ \protect{ 
Schematic side view of the geometry considered. The closed cylinder
$S'$ is made of the lateral surface $\Sigma$ and the
two cross sections perpendicular to the $Oz$ axis
($\Sigma_0$ and $\Sigma_L$}). The thick line denotes the
boundary $\partial {\cal R}$.
}
\end{figure}

%\end{multicols}


\begin{thebibliography}{99}
\bibitem[*]{umr1}
	Unit\'e Mixte de Recherche UMR 8627 du CNRS.
\bibitem[**]{umr2}
        Unit\'e Mixte de Recherche UMR 6619 du CNRS.
\bibitem{Larsen} 
        A.M. Larsen and D.G. Grier, Nature {\bf 385}, 230 (1997);
	J.C. Crocker and D.G. Grier, Phys. Rev. Lett. {\bf 77}, 1897 (1996).
\bibitem{Grier}
        D.G. Grier, Nature {\bf 393}, 621 (1998); 
\bibitem{dlvo} 
        E.J.W. Verwey and J.Th.G. Overbeek, {\it Theory of the Stability of
	Lyophobic Colloids\/} (Elsevier, New York, 1948).
\bibitem{Vonder} 
	K. Vondermasson, J. Bongers, A. Mueller and H. Versmold, 
	Langmuir {\bf 10}, 1351 (1994); 
	J.C. Crocker and D.G. Grier, Phys. Rev. Lett. {\bf 73}, 352 (1994).
\bibitem{Bowen}
        W.R. Bowen and A.O. Sharif, Nature {\bf 393}, 663 (1998).
\bibitem{Goulding}
        D. Goulding and J.P. Hansen, Europhys. Lett {\bf 46}, 407 (1999).
\bibitem{Neu}
        J.C. Neu, Phys. Rev. Lett. {\bf 82}, 1072 (1999).
\bibitem{Sader}
	J.E. Sader and D.Y. Chan, J. Colloid Interface Sci. {\bf 213}, 268 (1999).
\bibitem{Eigen}	
	Eigen and Wicke, J. Phys. Chem. {\bf 58}, 702 (1954).
\bibitem{Iglic}	
	V. Kralj-Igli$\check{\rm c}$ and A. Igli$\check{\rm c}$,
	J. Phys. II France {\bf 6}, 477 (1996).
\bibitem{Borukhov}
        I. Borukhov, D. Andelman and H. Orland, Phys. Rev. Lett. {\bf 79},
	435 (1997). 
\bibitem{Gouldingbis}	
        D. Goulding, to be published.
\bibitem{Stern} 
        O. Stern, Z. Elektrochem. {\bf 30}, 508 (1924). 
\bibitem{Kje} 	
	R. Kjellander,
	T. \AA kesson, B. J\"onsson and S. Mar$\check{\rm c}$ella, 
	J. Chem. Phys. {\bf 97},
	1424 (1992); B.W. Ninham and V.A. Parsegian, J. Theor. Biol. {\bf 31},
	408 (1971).
\bibitem{Biben} T. Biben, J.P. Hansen, Y. Rosenfeld, Phys. Rev. E {\bf 57}, 
        R3727 (1998).
	
	
	


\end{thebibliography}
\end{document}